\documentclass[a4paper,11pt]{article}
\pdfoutput=1
\usepackage{jinstpub}
\usepackage{indentfirst}
\usepackage{graphicx}
\usepackage{subfigure}
\usepackage{amsmath}
\usepackage{adjustbox}
\usepackage{multirow}

\title{Feasibility study of SiGHT: a novel ultra low background  photosensor for low temperature operation}
\author[a,b,c,h,1]{Y.~Wang,\note{Corresponding author.}}
\author[b,2]{A.~Fan,\note{Present address: SLAC National Accelerator Laboratory, 2575 Sand Hill Road, Menlo Park, CA 94205, USA.}}
\author[d]{G.~Fiorillo,}
\author[e]{C.~Galbiati,}
\author[a]{M.Y.~Guan,}
\author[f,g]{G.~Korga,}
\author[c]{E.~Pantic,}
\author[e,f]{A.~Razeto,}
\author[g]{A.~Renshaw,}
\author[d,e]{B.~Rossi,}
\author[b,f]{Y.~Suvorov,}
\author[b]{H.~Wang}
\author[a]{and C.G.~Yang}

\affiliation[a]{Key Laboratory of Particle Astrophysics, Institute of High Energy Physics, Chinese Academy of Sciences\\Beijing 100049, China}
\affiliation[b]{Department of Physics and Astronomy, University of California, Los Angeles\\California 90095, USA}
\affiliation[c]{Department of Physics, University of California, Davis\\California 95616, USA}
\affiliation[d]{Department of Physics, University ``Federico II'' \& INFN Napoli\\Via Cinthia, I-80126 Napoli, Italy}
\affiliation[e]{Department of Physics, Princeton University\\New Jersey 08544, USA}
\affiliation[f]{Laboratori Nazionali del Gran Sasso\\Assergi AQ 67010, Italy}
\affiliation[g]{Department of Physics, University of Houston\\Houston, Texas 77204, USA}
\affiliation[h]{School of Physical Sciences, University of Chinese Academy of Sciences\\Beijing 100049, China}

\emailAdd{ywangmax@physics.ucla.edu}

\abstract{Rare event search experiments, such as those searching for dark matter and observations of neutrinoless double beta decay, require ultra low levels of radioactive background for unmistakable identification. In order to reduce the radioactive background of detectors used in these types of event searches, low background photosensors are required, as the physical size of these detectors become increasing larger, and hence the number of such photosensors used also increases rapidly. Considering that most dark matter and neutrinoless double beta decay experiments are turning towards using noble liquids as the target choice, liquid xenon and liquid argon for instance, photosensors that can work well at cryogenic temperatures are required, 165 K and 87 K for liquid xenon and liquid argon, respectively.

The Silicon Geiger Hybrid Tube (SiGHT) is a novel photosensor designed specifically for use in ultra low background experiments operating at cryogenic temperatures. It is based on the proven photocathode plus silicon photomultiplier (SiPM) hybrid technology and consists of very few other, but also ultra radio-pure, materials like fused silica and silicon for the SiPM. The introduction of the SiGHT concept, as well as a feasibility study for its production, is reported in this paper.}


\begin{document}

\maketitle
\flushbottom

\section{Introduction and motivation}
Maintaining an ultra low radioactive background is currently the most essential requirement for rare event search experiments, such as those searching for dark matter and observations of neutrinoless double beta decay. Most of these experiments are using a noble liquid as the target, which requires the detector to be operated at cryogenic temperatures, 100\% of the time~\cite{liquidnoble}. For instance, liquid argon (boiling point at 87 K) is being used in the DarkSide-50 experiment~\cite{DS}, and liquid xenon (boiling point at 165 K) is being used in the XENON~\cite{XENON}, the LUX~\cite{LUX}, the PandaX~\cite{PandaX} and the EXO~\cite{EXO} experiments. In order to maximize the light yield of the detector, improvement of the photosensor performance becomes a necessary challenge to be overcome for the next generation of detectors. Besides the need for excellent and long term maintained performance at cryogenic temperatures, the radioactivity of the photosenors themselves also begin to dominate the radioactive background inside the detectors. Considering that the target mass of upcoming detectors will be on the scale of tonnes to tens of tonnes~\cite{tonscale}, the accumulation of radioactivity from the large amount of photosensors will begin to overwhelm the detector and effect performance, and hence the potential physics reach of said experiments.

The photosensors being used in the current generation of detectors being operated and constructed consist of mainly photomultiplier tubes (PMTs). The most popular cryogenic PMTs are produced by Hamamatsu, which implement the new technique of using a bi-alklai photocathode with quantum efficiency (QE) that can reach $\sim$30\% and a metal-based cryogenic vacuum seal. The Hamamatsu R11065 and Hamamatsu R11410 PMTs are currently used in the DarkSide-50 and the XENON-1T experiments~\cite{R11410}, respectively. Although the performance of such PMTs are improved at cryogenic temperatures, the conventional multi-dynode electron amplification structure still contains too many materials and is the main contribution of radioactive background of an entire PMT. In order to further reduce the background from the photoelectronics, the conventional multi-dynode structure can be replaced by new electronic devices, such as avanlache photodiodes (APDs) and silicon photomultipliers (SiPMs). APDs and SiPMs are entirely made of silicon, so no matter the quantity, they are intrinsically more pure and more efficient than the multi-dynode structure of a traditional PMT. The QUartz Photon Intensifying Detector(QUPID) is a photosensor using the photocathode and APD hybrid technologies, designed for operation in liquid xenon, with a prototype already produced by Hamamatsu and tested at the UCLA dark matter lab~\cite{QUPIDa,QUPIDe}. For some reasons, this project has been terminated by Hamamatsu in 2012.

The Silicon Geiger Hybrid Tube (SiGHT) is a novel hybrid photosensor that has been proposed to be developed for the use in future ultra low background experiments~\cite{SiGHT}. It will use a combination of a photocathode and a SiPM, enabling a lower operating voltage and bias voltage compare to the QUPID, while maintaining good resolution and mechanical characteristics in both liquid xenon and liquid argon. To intrinsically minimize the radioactive background, the entire body of SiGHT is made of ultra pure fused silica, with no other components aside from one SiPM and two super invar electrodes. All the electrical connections are achieved by thin film deposition, with thickness precisely controlled to minimize the total activity of materials. As expected, the estimated radioactivity of one 3" diameter SiGHT photosensor is less than 0.04 mBq, while the QE is larger than 30\% at cryogenic temperatures, for the desired wavelength of light. Currently the first SiGHT prototype is under development at the UCLA SiGHT lab. Some of the related techniques to produce the photosensor are described in the following sections.

\section{SiGHT design and field simulation}

\begin{table}[ht]
\centering
\caption{Intrinsic radioactivity of a single 3" diameter SiGHT. The masses are measured from the SiGHT model. The radioactivity of each component is estimated by the related references.}
\label{materials}
\begin{adjustbox}{width=1\textwidth}
\begin{tabular}{c c c c c c c c c}
\hline
\hline
\multirow{2}{*}{Material} & Mass & $^{238}$U & $^{232}$Th & $^{60}$Co & $^{40}$K & Total & Activity per & \multirow{2}{*}{Reference} \\
 & [g] & [mBq/kg] & [mBq/kg] & [mBq/kg] & [mBq/kg] & [mBq/kg] & SiGHT [mBq] & \\
\hline
Fused silica & 109 & 0.008 & 0.01 & - & - & 0.018 & 0.002 & DarkSide\footnotemark[3] \\
Super invar\footnotemark[4] & 0.13 & <18.4 & <0.72 & <0.15 & <1.98 & <21.25 & <0.003 & ~\cite{XENON1TPMT} \\
SiPM & 0.026 & <0.025 & <0.003 & - & - & <0.028 & <7.3$\times$10$^{-7}$ & ~\cite{SiPMBG} \\
Indium & 0.035 & <720 & <3.1 & <0.2 & <9.3 & <732.6 & <0.03 & DarkSide \\
Copper & 0.066 & <0.06 & <0.02 & - & 0.12 & <0.2 & <1.4$\times$10$^{-5}$ & DarkSide \\
Chromium & 0.007 & <61.73 & <81.4 & - & <216.72 & <359.85 & <2.52$\times$10$^{-3}$ & ~\cite{CrBG} \\
Aluminum & 0.003 & <0.055 & <0.002 & <2.4$\times$10$^{-4}$ & <0.011 & <0.07 & <2.1$\times$10$^{-7}$ & ~\cite{XENON1TPMT} \\
\hline
Total & 109.267 &  &  &  &  &  & <0.04 & \\
\hline
\hline
\end{tabular}
\end{adjustbox}
\end{table}

\footnotetext[3]{The activity is measured by DarkSide collaboration.}
\footnotetext[4]{The activity is estimated by the measurement of Co-free kovar.}

\begin{figure}[ht]
\centering 
\subfigure[3D view]{ \label{fig:SiGHT_view:a} 
\includegraphics[width=7cm]{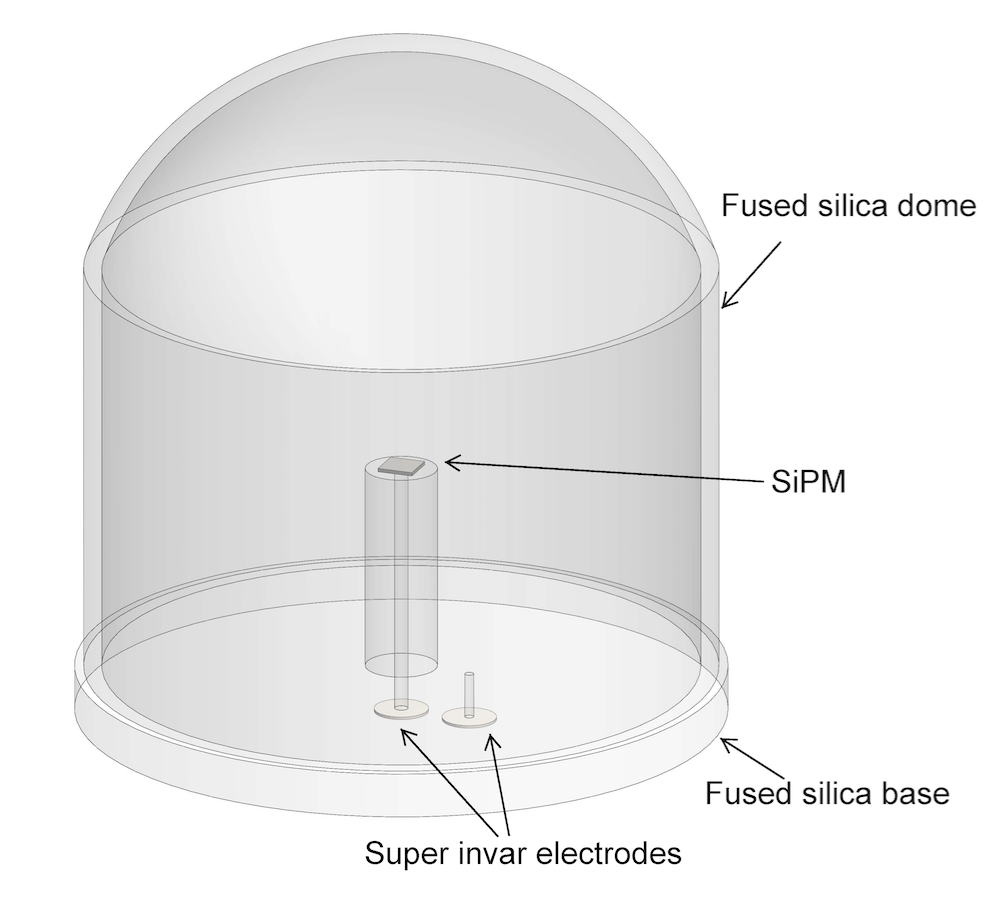}} 
\hspace{0.5in} 
\subfigure[SiGHT model]{ \label{fig:SiGHT_view:b} 
\includegraphics[width=5cm]{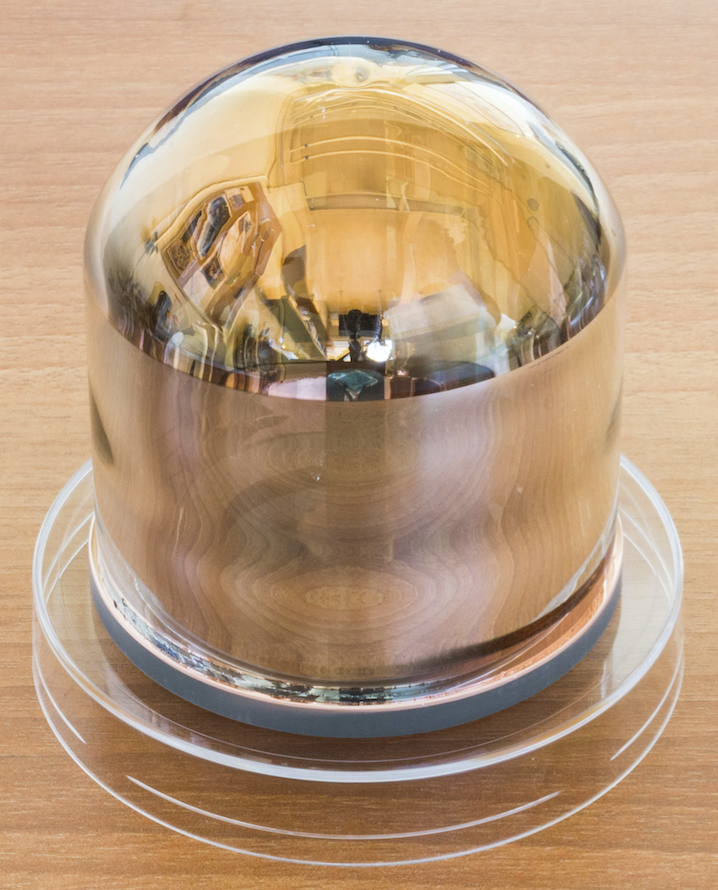}} 
\caption{$\emph{Left}$: 3D view of SiGHT, SiGHT only consists of one fused silica dome, one fused silica base, one SiPM and two super invar electrodes; $\emph{Right}$: Picture of a SiGHT model fabricated and assembled at the UCLA SiGHT lab.} 
\label{fig:SiGHT_view} 
\end{figure}

Figure~\ref{fig:SiGHT_view} shows a 3D view and a picture of a 3" diameter SiGHT. The body of a SiGHT consists of two fused silica parts, the dome and the base. One $3\times3$ mm$^{2}$ SiPM is fixed to the top of the fused silica pillar, which is part of the base, and two 8 mm diameter super invar discs are affixed to the bottom of the base to function as electrodes. The electrical connections inside SiGHT are achieved by different kinds
of thin film metal layers. Table~\ref{materials} shows the content of materials used to build a 3" diameter SiGHT, as well as their radioactivities. According to the estimation, the overall radioactivity of a single 3" diameter SiGHT can be maintained below 0.04 mBq.

Once proper operating voltages are applied to the thin film metal layers and the SiPM, photoelectrons are generated from the coming photons when they hit the photocahtode, after traveling through the fused silica of the dome. The electric field that is maintained inside the SiGHT evacuated dome will allow photoelectrons to drift and be focused onto the SiPM for signal readout. SiGHT is considered to be a hybrid design because of the use of the combination of the photocathode and the SiPM, which could be used to detect light on its own. However, the use of the photocathode allows for maximization of light sensitive area in a detector, while maintaining a lower number of required readout electronics.

\begin{figure}[ht]
  \centering
  \includegraphics[width=12cm]{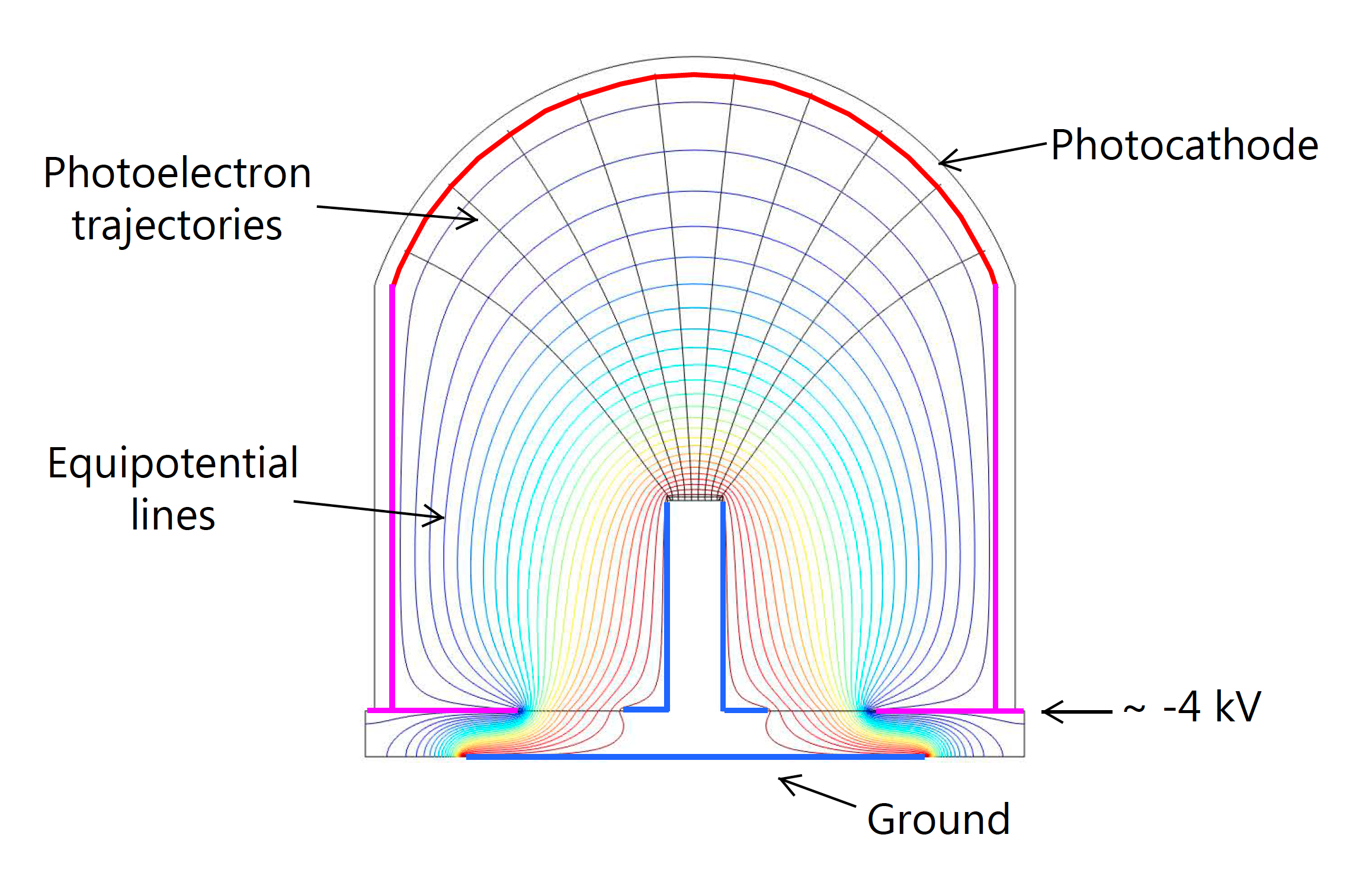}
  \caption{Simulation of the electric field and electron flight trajectories for SiGHT, done with COMSOL Multiphysics. $-$4 kV is applied to the photocathode, while the blue lines represent ground potential.}
  \label{fig:SiGHT_field}
\end{figure}

In order to maximize the photoelectron collection efficiency, an electric field simulation was done to optimize the geometry of SiGHT. Figure~\ref{fig:SiGHT_field} shows the resulting electric field and electron flight trajectories from a finite element analysis (FEA) done with the COMSOL Multiphysics software~\cite{comsol}, and based upon the optimized geometry. Ground potential is held at the blue lines in the figure, while $-$4 kV operating voltage is applied to the photocathode (red arc in the figure) through conductive thin film metal layers (purple lines in the figure). From the photocathode to the SiPM, the uniformity of the electric field is well indicated by the equipotential line distribution. Ten electrons were uniformly generated across the photocathode and they all move towards the SiPM due to the supplied electric field. From the simulation results, even electrons produced at the edge of the photocathode can make their way to the SiPM, allowing for almost 100\% photoelectron collection efficiency onto the SiPM (does not take into account fill factor of the SiPM). This being the case, the photon detection effiency (PDE) of SiGHT is basically determined by the photocathode QE and the SiPM electron detection efficiency (EDE) and is given as,
\begin{equation}
\label{eq:eff}
PDE_{SiGHT}=QE_{photocathode}\cdot EDE_{SiPM}.
\end{equation}

\section{Feasibility study}
Before the first prototype is assembled, several techniques related to SiGHT development were studied. According to the required performance of SiGHT, there are three important factors which need to be taken into account: maintaining a vacuum seal at cryogenic temperatures, a photocathode which is efficient at cryogenic temperatures, and performance of the SiPM. These three factors will be detailed in the following sections.

\subsection{Cryogenic vacuum seal}
Maintaining a good vacuum seal at cryogenic temperatures, while using the minimum quantity of material is the goal in mind for SiGHT. Since SiGHT is mostly composed of fused silica, an indium seal should function as the best solution. Indium seals are widely used in cryogenic systems and can be easily achieved by using thin film evaporation technology to minimize the activity~\cite{indium}. As far as the type of material is concerned, there are two kinds of seals within the SiGHT design: 1) the seal between the dome and the base, and 2) the seal between the base and the two super invar electrodes.

\begin{figure}[ht]
  \centering
  \includegraphics[width=12cm]{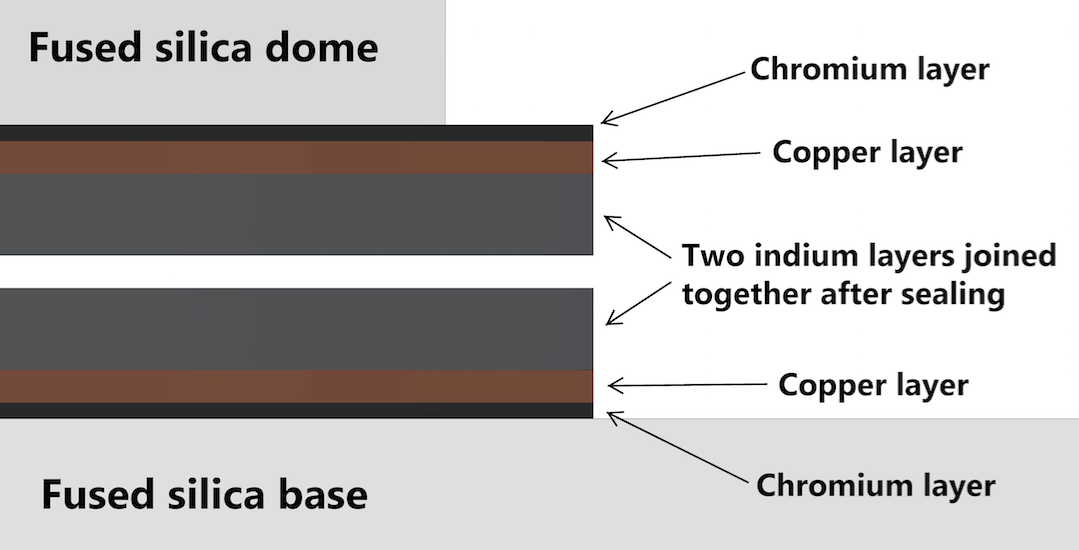}
  \caption{Sketch of the indium seal between the fused silica dome and the fused silica base. 100 nm of chromium and 200 nm of copper are coated as under layers. The two indium layers join together after sealing to maintain the vacuum inside SiGHT.}
  \label{fig:Indium_seal}
\end{figure}

An indium seal between the dome and the base is easily accomplished, because the dome and the base are made of the same material and thus will have the same thermal expansion. In order to adhere an indium layer on fused silica closely, chromium and copper under layers are required. Chromium is a metal which can adhere to fused silica very strongly, and copper can adhere to both indium and chromium very well. After several tests, the optimal solution was found. Starting with the fused silica, from bottom to top, the order of layers is: 100 nm of chromium, 200 nm of copper and then several microns of indium, see figure~\ref{fig:Indium_seal}. The thermal resistance evaporation method is used for the thin film metal layers deposition. A test pair of fused silica disks and tubes were first coated with chromium, copper and indium and then placed on a hot plate held at 200$^\circ \mathrm{C}$ inside a vacuum chamber for 24 hours, to melt the indium completely and form the seal. Once the seal is complete, a leak test is made using an Aligent leak detector to confirm that the helium leak rate was lower than 1$\times$10$^{-11}$ atm-cc/sec, indicating that the seal is more than enough to maintain the vacuum inside SiGHT. Then the sealed parts were dropped into liquid nitrogen, then removed and allowed to warm to room temperature, and repeated 4 times before a second leak test was done. The helium leak rate during the second test was still lower than 1$\times$10$^{-11}$ atm-cc/sec, proving that this seal can survive environments at cryogenic temperatures.

The indium seal between the fused silica base and the two super invar electrodes is almost identical to the seal between the dome and the base. The thermal expansion of super invar is similar to the thermal expansion of fused silica~\cite{Invar}. The only difference between this seal and the previously discussed seal, is there is no need for chromium and copper under layers on the super invar, since the indium can directly adhere to it. The same procedure of leak testing was done for the seal between the base and the super invar electrodes, with the helium leak rate always lower than 1$\times$10$^{-11}$ atm-cc/sec.

\subsection{Photocathode}
The photocathode is one of the two parts which determines the PDE of SiGHT. Low temperature rare event detections require a photocathode with as high a QE as possible, a low dark count rate and good linearity. QE is mostly relative to the wavelength of light. The wavelength of light emitted from argon and xenon scintillation is 128 nm and 178 nm, respectively~\cite{wavelength}. For argon, since 128 nm light does not transmit through the fused silica dome, a wavelength shifter tetraphenyl butadiene (TPB) has to be used to shift the wavelength from 128nm to a spectrum which peaks at 420 nm~\cite{TPB}. Thus, the QE for SiGHT will be studied for 178 nm and 420 nm wavelength light specifically.

\begin{figure}[ht]
  \centering
  \includegraphics[width=12cm]{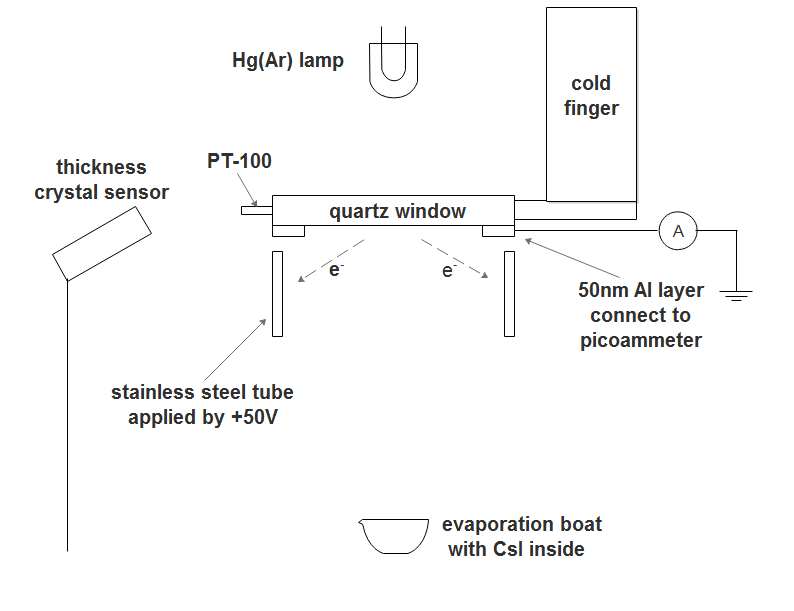}
  \caption{CsI photocathode cryogenic test setup. A 3" diameter quartz window with a 50 nm aluminum ring coating is placed on a copper plate connected to a liquid nitrogen cold finger. The aluminum ring is virtually connected to ground through a picoammeter, and below this there is a stainless steel tube with 50 V applied to it to pull the photoelectrons from the photocathode. A PT-100 is positioned next to the quartz window to monitor the temperature. A CsI thermal evaporation system is located on the bottom of setup.}
  \label{fig:PC_setup}
\end{figure}

\begin{figure}[ht]
  \centering
  \includegraphics[width=12cm]{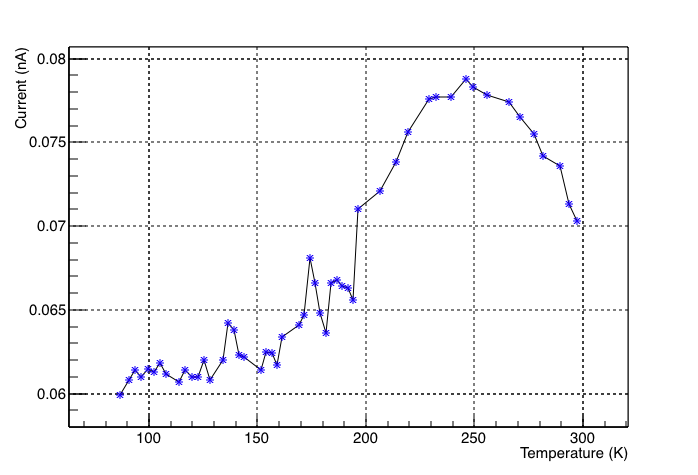}
  \caption{Current as a function of temperature for the CsI photocathode. As the temperature goes down, the current increases until 245 K, at which point it begins to drop.}
  \label{fig:PC_current}
\end{figure}

Hamamatsu has developed and successfully produced an ultra bi-alkali photocathode for low temperature operation. It has $\sim$43\% QE at peak wavelength and good linearity at liquid argon temperature~\cite{PC}. SiGHT will eventually use bi-alkali for the photocathode, which will consist of an alkali antimonide having a composition of K-Sb-Cs, and a layer of transparent electrical conductor to be coated above the photocathode to increase its electrical conductivity. Even though the transparency will become worse because of the conductive layer, the increase in electrical conductivity will allow for a higher overall QE.

The first SiGHT prototype used cesium iodide (CsI) as the photocathode, since it can be easily produced in lab~\cite{CsI}. In order to understand the cryogenic performance of the CsI photocathode, a 3" diameter quartz window coated with a 50 nm aluminum ring is placed in the setup shown in figure~\ref{fig:PC_setup}, in order to carry out the test at cryogenic temperatures. The quartz window is virtually connected to ground through a picoammeter which can measure the leak current. A Hg (Ar) lamp with peak wavelength at 254 nm flashes above the window so that photoelectrons can be extracted from the plate by a stainless steel tube below the window, held at 50 V. A CsI source is located in a tungsten boat at the bottom of chamber. The evaporation rate is monitored by a crystal sensor standing next to the window. 

After 24 hours pumping and baking, a thin layer of CsI is coated onto the quartz window by monitoring the current with the picoammeter. Once the highest QE is achieved, coating is stopped manually. A nitrogen cooling cold finger cools down the window after the entire system gets back to room temperature. The current is monitored while cooling from room temperature to 87 K.

Figure~\ref{fig:PC_current} shows the current of the CsI photocahtode as a function of temperature. At the beginning, the current is increasing some as the temperature goes down. This is because the lower temperature reduces the work function of electrons, making the photoelectrons easier to escape than at room temperature. As further decrease of temperature occurs (less than 245 K), the electrical conductivity of CsI becomes worse and gradually prevents the supply of the escaped electrons from ground. This causes the current to slightly go down and eventually drop to 0.06 nA at 87 K. In conclusion, while the current dropped slightly with decreasing temperature, the ability for photon conversion to happen still remained.

\subsection{SiPM}
Although SiPMs are mainly designed for photon collection, it is essentially a geiger mode APD (G-APD) matrix which is composed of 100 to 10000 APDs~\cite{SiPM}. In principle, it can be used as an electron counter, since the pixel is fired by energy deposition. Considering the difference between photons and electrons, the SiPM for SiGHT should be bare, which means no protection layer to prevent electrons entering the p-n junctions. In the meanwhile, low dark count rate, high gain and single photoelectron detection ability are required.

\begin{figure}[ht]
  \centering
  \includegraphics[width=12cm]{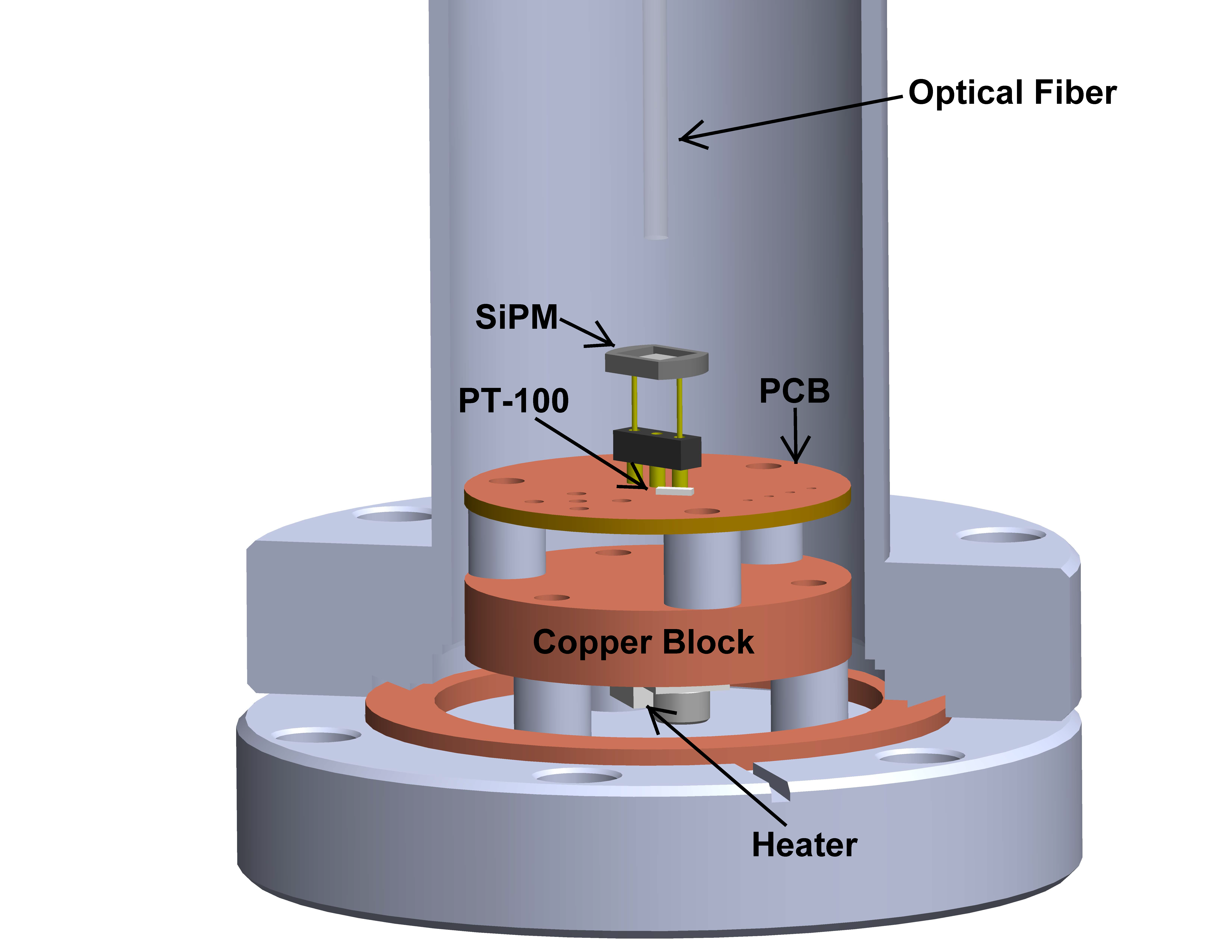}
  \caption{The setup used for the laser pulsed SiPM cryogenic test. SiPM is fixed onto a PCB sealed in a vacuum chamber. The temperature can be precisely controlled by a LakeShore 325 temperature controller connected to a PT-100 and a heater that are placed next to and below SiPM, respectively. The laser light is coupled into an optical fiber and fed through into the vacuum chamber using a custom made feedthrough. The entire chamber is dropped into a liquid nitrogen bath for cooling.}
  \label{fig:SiPM_setup}
\end{figure}

A SensL C30035 SiPM~\cite{SensL} was used for the first cryogenic test, the setup of which can be seen in figure~\ref{fig:SiPM_setup}. A SiPM is fixed on a PCB with a PT-100 temperature sensor near it. The entire setup is sealed in a small vacuum chamber and dropped into a liquid nitrogen bath for cooling. The temperature is controlled via a LakeShore 325 temperature controller by adjusting the voltage applied to the heater that is placed on the back side of copper block. An operational amplifier is connected to the SiPM outside of the vacuum chamber to amplify the signal before being readout using a LeCroy oscilloscope. Several different temperature points were measured, the results are given in the following.

\begin{figure}[ht]
  \centering
  \includegraphics[width=12cm]{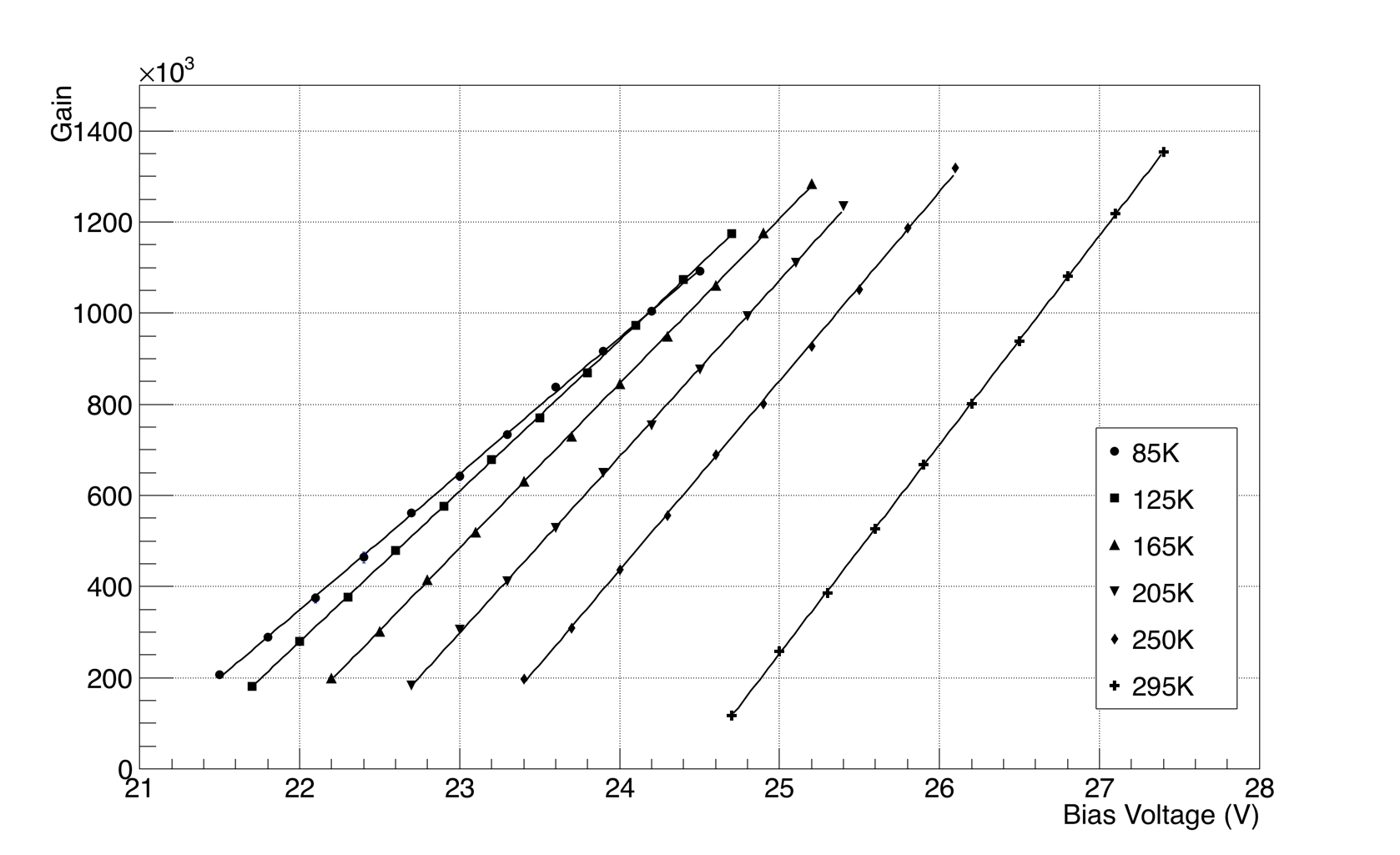}
  \caption{SensL C30035 SiPM gain versus bias at different temperatures. By adjusting the bias voltage the gain can easily reach 10$^{6}$.}
  \label{fig:Gain_vs_bias}
\end{figure}

Figure~\ref{fig:Gain_vs_bias} shows the gain of the SiPM as a function of bias voltage, for different temperatures. As the temperature decreases, the bias voltage that can offer the same gain decreases because the breakdown voltage of the G-APD decreases. Whether in liquid argon (87 K) or liquid xenon (165 K), the gain of the SiPM can reach 10$^{6}$ by adjusting the bias voltage.

\begin{figure}[ht]
  \centering
  \includegraphics[width=12cm]{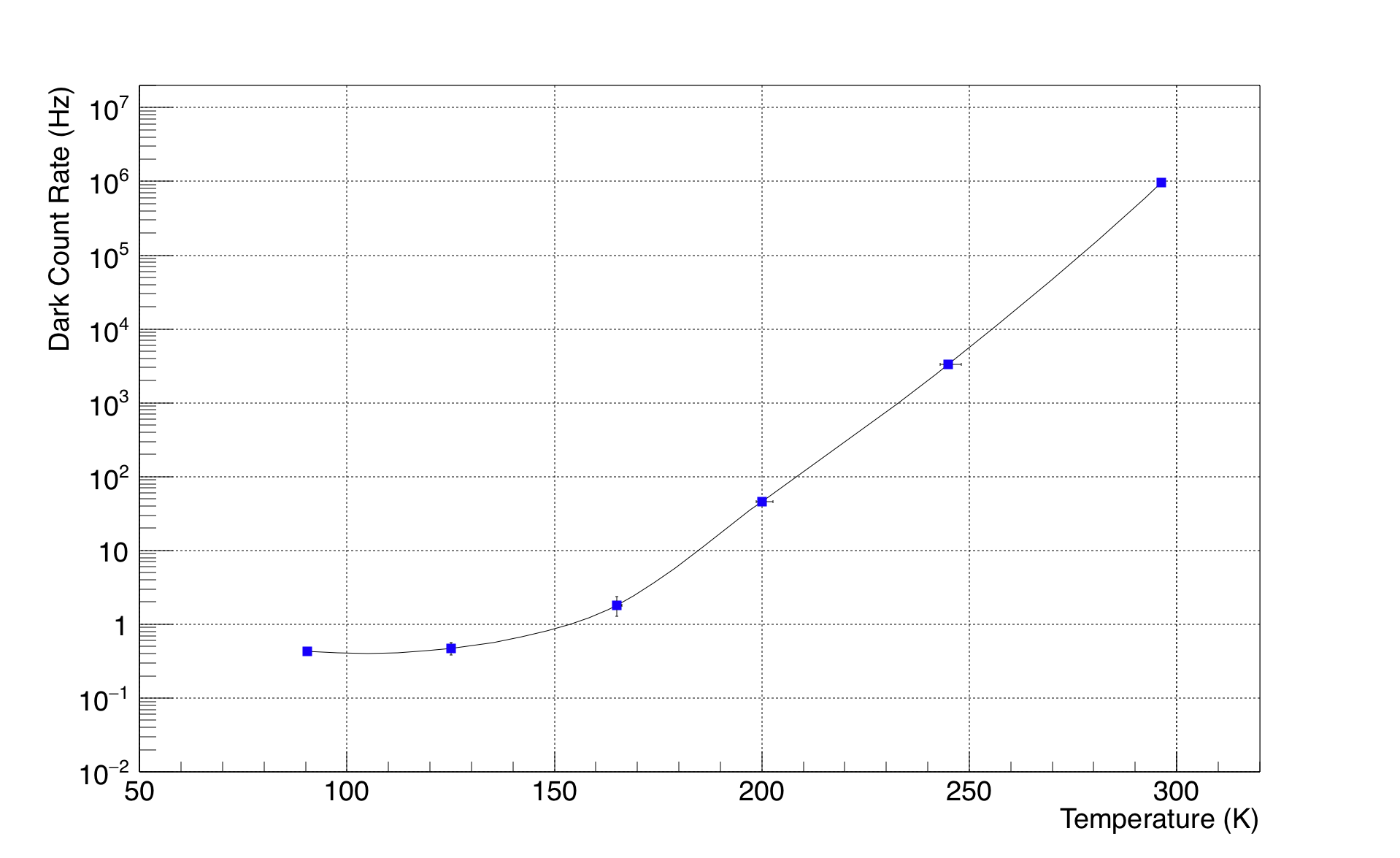}
  \caption{SensL C30035 SiPM dark count rate as a function of temperature. During the entire measurement, the gain is fixed as 10$^{6}$ by adjusting the bias voltage at different temperature points. At room temperature (296 K), the dark count rate is $\sim$1 MHz; at liquid xenon temperature (165 K), the dark count rate is $\sim$2 Hz; at liquid argon temperate (87 K), the dark count rate is $\sim$0.5 Hz.}
  \label{fig:DCR_Temp}
\end{figure}

The dark count rate of the SiPM decreases dramatically as the temperature is decreased. As shown in figure~\ref{fig:DCR_Temp}, the dark count rate at 87 K is lower than 0.1 Hz/mm$^{2}$, which is six orders of magnitude lower than the value measured at room temperature. The dark count rates at 87 K and 165 K are low enough to be accepted by the performance of SiGHT.

\begin{figure}[ht]
  \centering
  \includegraphics[width=12cm]{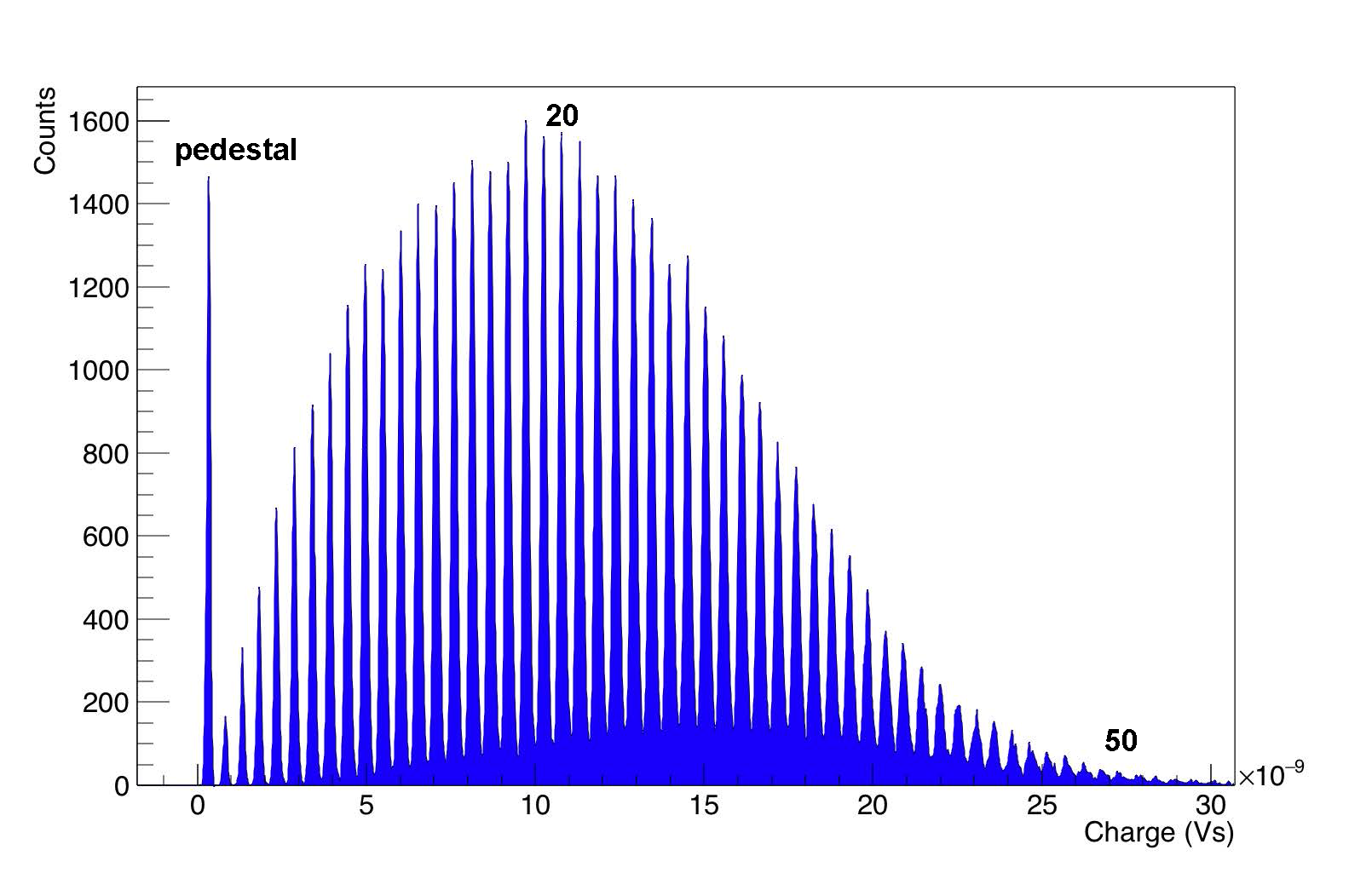}
  \caption{SensL C30035 SiPM photon counting capability at 87 K.}
  \label{fig:counting}
\end{figure}

\begin{figure}[ht]
  \centering
  \includegraphics[width=12cm]{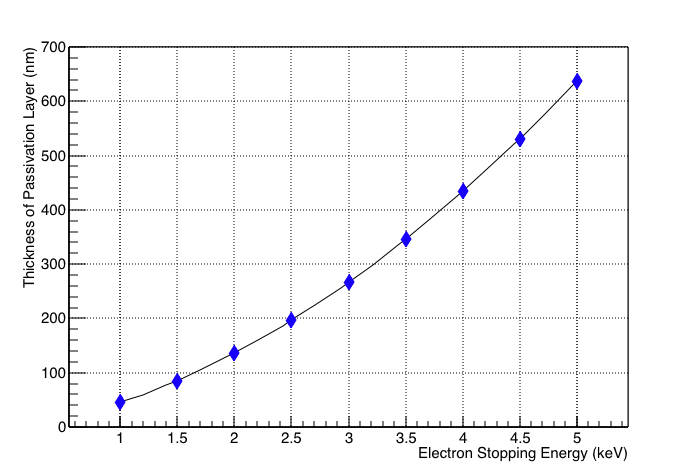}
  \caption{Electron stopping energy in SiO$_{2}$}
  \label{fig:stopping_energy}
\end{figure}

With the photon flux produced by the laser, very high resolution photon counting capability of the SiPM can be measured and is shown in figure~\ref{fig:counting}. Up to 50 photons can be detected at the same time and it is believed even more can be counted since the true limitation here may be the charge amplifier and oscilloscope readout. 

With regards to the photoelectron detection efficiency, the most critical issue is to prove the electron detection ability of the SiPM. Since each pixel of the SiPM has a SiO$_{2}$ passivation layer in the front, there is an energy threshold for the photoelectron to make it through that layer and eventually deposit energy in a G-APD. Figure~\ref{fig:stopping_energy} shows the electron stopping energy as a function of the thickness of the SiO$_{2}$ passivation layer calculated with NIST electron stopping power~\cite{NIST}. In order to minimize the operating voltage, the SiPM passivaiton layer should be made as thin as possible. Another effective way to let electrons be detected by SiPM easier, is using back illuminated SiPMs~\cite{BIDSiPM} instead of front illuminated SiPMs. This is because the back illuminated SiPMs move all the metal wirings below the actual photodiode so electrons do not have to go through metal wirings before depositing energy inside the photodiode.

\section{Conclusion and discussion}
Base on the feasibility studies described above, the development of SiGHT, a novel hybrid ultra low background photosensor is viable, and will be pursued based on it being an essential components of future large scale rare event search experiments. Compared to conventional PMTs, SiGHT has extremely low radioactive background and stable working performance at cryogenic temperatures.

The first SiGHT prototype development is in progress at the UCLA SiGHT lab, and is proposed to use CsI as the photocathode and a SiPM as the photoelectron counter. Most of the relative techniques have already been tested and confirmed to be suitable for this application. Furthermore, a test will be carried out in which electrons will be directly focused onto a SiPM in order to study the detailed electron detection efficiency. Finally, a new method for bi-alkali photocathode production is being studied in parallel, and will eventually replace the CsI photocathode to be used initially. The readout electronics will be developed with the help of the DarkSide collaboration.

The expected characteristics of SiGHT would make it an ideal photosenor to replace PMTs in tonne scale and beyond noble liquid detectors used for rare event searches. In particular, it will serve as an important component for future direct dark matter search and neutrinoless double beta decay observation experiments. 

\acknowledgments

This work is supported by NSF PHY-1314501, PHY-1413358, PHY-1455351 as well as the Institute of High Energy Physics, Chinese Academy of Sciences.

The authors also thank Artin Teymourian and Alexey Lyashenko, former researchers at UCLA who did a lot of preparatory work for the SiGHT development. And last but not least, the authors thank the UCLA physics department machine shop, headed by Harry Lockart, and did countless work for the fabrication of the experimental setup.

\end{document}